\documentclass[twocolumn,showpacs,preprintnumbers,amsmath,amssymb]{revtex4}
\usepackage{graphicx}
\usepackage{dcolumn}
\usepackage{bm}
\begin{document}
\title{Direct measurements of the spin and the cyclotron gaps\\ in a
2D electron system in silicon}
\author{V.~S. Khrapai, A.~A. Shashkin, and V.~T. Dolgopolov}
\affiliation{Institute of Solid State Physics, Chernogolovka, Moscow
District 142432, Russia}
\begin{abstract}
Using magnetocapacitance data in tilted magnetic fields, we directly
determine the chemical potential jump in a strongly correlated
two-dimensional electron system in silicon when the filling factor
traverses the spin and the cyclotron gaps. The data yield an
effective $g$ factor that is close to its value in bulk silicon and
does not depend on filling factor. The cyclotron splitting
corresponds to the effective mass that is strongly enhanced at low
electron densities.
\end{abstract}
\maketitle

A two-dimensional (2D) electron system in silicon
metal-oxide-semiconductor field-effect transistors (MOSFETs) is
remarkable due to strong electron-electron interactions. The Coulomb
energy overpowers both the Fermi energy and the cyclotron energy in
accessible magnetic fields. The Landau-level-based considerations of
many-body gaps \cite{ando,yang}, which are valid in the weakly
interacting limit, cannot be directly applied to this strongly
correlated electron system. In a perpendicular magnetic field, the
gaps for charge-carrying excitations in the spectrum should originate
from cyclotron, spin, and valley splittings and be related to a
change of at least one of the following quantum numbers: Landau
level, spin, and valley indices. However, the gap correspondence to a
particular single-particle splitting is not obvious \cite{brener},
and the origin of the excitations is unclear. In a recent theory
\cite{iordan}, the strongly interacting limit has been studied, and
it has been predicted that in contrast to the single-particle
picture, the many-body gap to create a charge-carrying (iso)spin
texture excitation at integer filling factor is determined by the
cyclotron energy. This is also in contrast to the square-root
magnetic field dependence of the gap expected in the weakly
interacting limit \cite{ando,yang}.

A standard experimental method for determining the gap value in the
spectrum of the 2D electron system in a quantizing magnetic field is
activation energy measurements at the minima of the longitudinal
resistance \cite{englert,klein,dol88,usher}. Its disadvantage is that
it yields a mobility gap which may be different from the gap in the
spectrum. In Si MOSFETs, the activation energy as a function of
magnetic field was reported to be close to half of the
single-particle cyclotron energy for filling factor $\nu=4$, while
decreasing progressively for the higher $\nu$ cyclotron gaps
\cite{englert,klein,dol88}. At low electron densities, an interplay
was observed between the cyclotron and the spin gaps, manifested by
the disappearance of the cyclotron ($\nu=4$, 8, and 12) minima of the
longitudinal resistance \cite{krav00}. On the contrary, for the 2D
electrons in GaAs/AlGaAs heterostructures, the activation energy at
$\nu=2$ exceeded half the single-particle cyclotron energy by about
40\% \cite{usher}. Another, direct method for determining the gap in
the spectrum is measurement of the chemical potential jump across the
gap \cite{smith85,aristov,valley}. It was applied to the 2D electrons
in GaAs \cite{smith85} and gave cyclotron gap values corresponding to
the band electron mass \cite{aristov}. Recently, the method has been
used to study the valley gap at the lowest filling factors in the 2D
electron system in silicon which has been found to be strongly
enhanced and increase linearly with magnetic field \cite{valley,rem}.

The effective electron mass, $m$, and $g$ factor in Si MOSFETs have
been determined lately from measurements of the parallel magnetic
field of full spin polarization in this electron system and of the
slope of the metallic temperature dependence of the conductivity in
zero magnetic field \cite{gm}. It is striking that the effective mass
becomes strongly enhanced with decreasing electron density, $n_s$,
while the $g$ factor remains nearly constant and close to its value
in bulk silicon. This result is consistent with accurate measurements
of $m$ at low $n_s$ by analyzing the temperature dependence of the
Shubnikov-de~Haas oscillations in weak magnetic fields in the
low-temperature limit \cite{us,rem1}. A priori it is unknown whether
or not the so-determined values $g$ and $m$ correspond to the spin
and the cyclotron splittings in strong perpendicular magnetic fields.

In this paper, we report the first measurements of the chemical
potential jump across the spin and the cyclotron gaps in a 2D
electron system in silicon in tilted magnetic fields using a
magnetocapacitance technique. We find that (i) the $g$ factor is 
close to its value in bulk silicon and does not change with filling 
factor, in contrast to the strong dependence of the valley gap on 
$\nu$; and (ii) the cyclotron splitting is determined by the 
effective mass that is strongly enhanced at low electron densities. 
We also verify the systematics of the gaps in that the measured 
$\nu=4$, 8, and 12 cyclotron gap decreases with parallel magnetic 
field component by the same amount as the $\nu=2$, 6, and 10 spin gap 
increases.

Measurements were made in an Oxford dilution refrigerator with a
base temperature of $\approx 30$~mK on high-mobility (100)-silicon
MOSFETs (with a peak mobility close to 2~m$^2$/Vs at 4.2~K) having
the Corbino geometry with diameters 250 and 660~$\mu$m. The gate
voltage was modulated with a small ac voltage 15~mV at frequencies
in the range 2.5 -- 25~Hz and the imaginary current component was
measured with high precision using a current-voltage converter and
a lock-in amplifier. Care was taken to reach the low frequency
limit where the magnetocapacitance, $C(B)$, is not distorted by
lateral transport effects. A dip in the magnetocapacitance at
integer filling factor is directly related to a jump, $\Delta$, of
the chemical potential across a corresponding gap in the spectrum
of the 2D electron system, and therefore we determine $\Delta$ by
integrating $C(B)$ over the dip in the low temperature limit where
the magnetocapacitance saturates and becomes independent of
temperature \cite{valley}.

\begin{figure}\vspace{2mm}
\scalebox{0.45}{\includegraphics[clip]{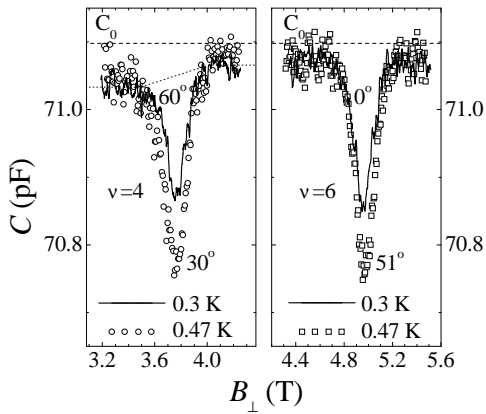}}
\caption{\label{fig1} Magnetocapacitance traces for $n_s=3.65\times
10^{11}$~cm$^{-2}$ (left panel) and $n_s=7.21\times
10^{11}$~cm$^{-2}$ (right panel). Also shown are the geometric
capacitance $C_0$ (dashed lines) and the step function
$C_{\text{ref}}$ (dotted line).}\vspace{-0.1in}
\end{figure}

Typical magnetocapacitance traces taken at different electron
densities, temperatures, and tilt angles are displayed in
Fig.~\ref{fig1} near the filling factor $\nu=hcn_s/eB_\perp=4$ and
$\nu=6$. The magnetocapacitance shows narrow minima at integer $\nu$
which are separated by broad maxima, the oscillation pattern
reflecting the modulation of the thermodynamic density of states,
$D$, in quantizing magnetic fields: $1/C=1/C_0+1/Ae^2D$ (where $C_0$
is the geometric capacitance between the gate and the 2D electrons,
and $A$ is the sample area) \cite{smith85}. As the magnetic field is
increased, the maximum $C$ approaches the geometric capacitance
indicated by the dashed lines in Fig.~\ref{fig1}. Since the
magnetocapacitance $C(B)<C_0$ around each maximum is almost
independent of magnetic field, this results in asymmetric minima of
$C(B)$, the asymmetry being more pronounced for $\nu=4$, 8, and 12.
The chemical potential jump at integer $\nu=\nu_0$ is determined by
the area of the dip in $C(B)$:

\begin{equation}
\Delta=\frac{Ae^3\nu_0}{hcC_0}\int_{\text{dip}}\frac{C_{\text{ref}}-
C}{C}dB_\perp, \label{Delta}\end{equation}
where $C_{\text{ref}}$ is a step function that is defined by two
reference levels corresponding to the capacitance values at the low
and high field edges of the dip as shown by the dotted line in
Fig.~\ref{fig1}. The so-determined $\Delta$ is smaller than the level
splitting by the level width. The last is extracted from the data by
substituting $(C_0-C_{\text{ref}})/C$ for the integrand in
Eq.~(\ref{Delta}) and integrating for the case of resolved levels
between the magnetic fields $B_1=hcn_s/e(\nu_0+1/2)$ and
$B_2=hcn_s/e(\nu_0-1/2)$.

\begin{figure}\vspace{2mm}
\scalebox{0.45}{\includegraphics[clip]{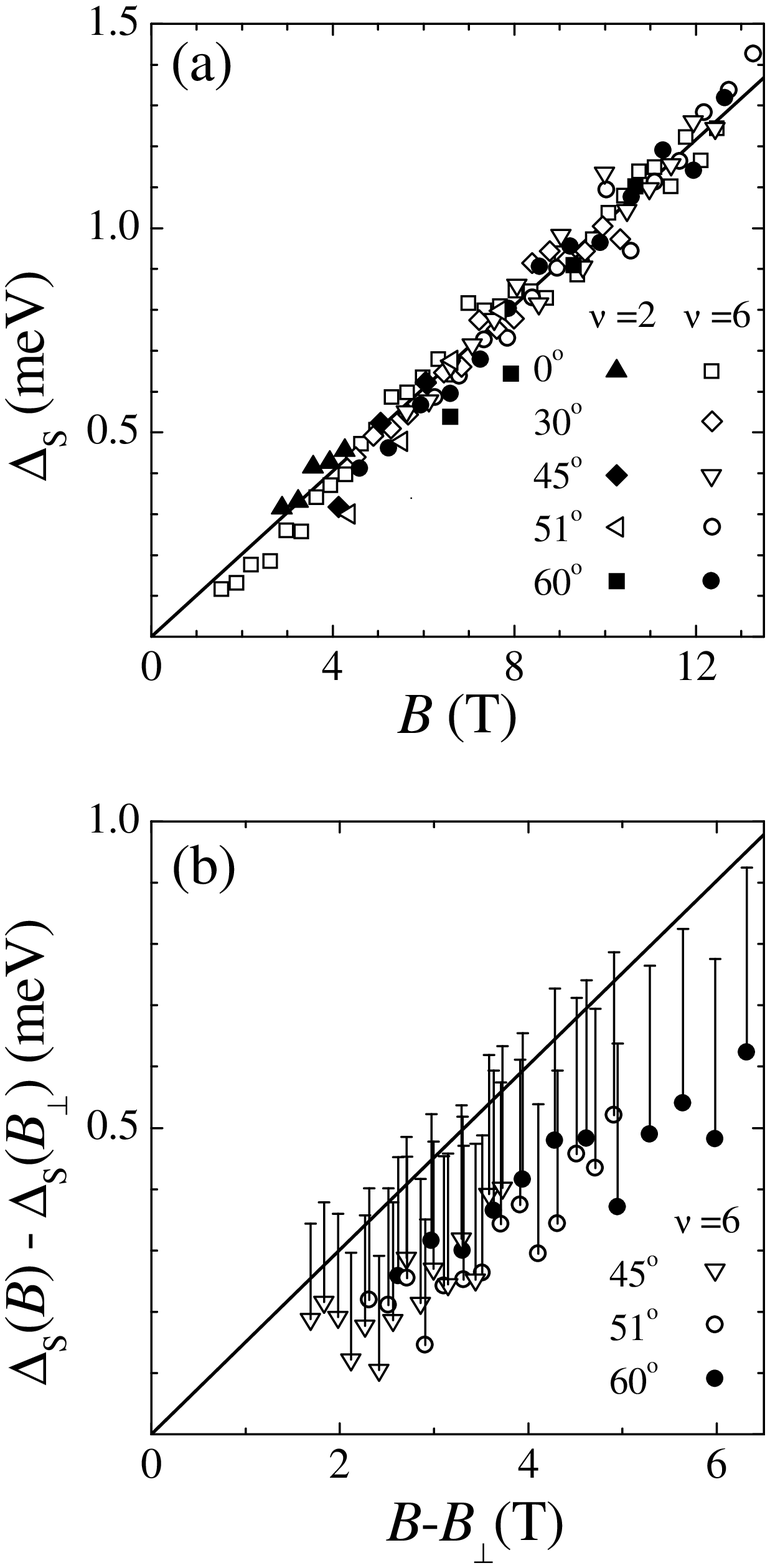}}
\caption{\label{fig2} (a) Chemical potential jump across the spin gap
as a function of magnetic field. The slope of the solid line gives a
lower boundary for $g\approx 1.75$. (b) Change of the spin gap with
$B_\parallel$ at different values of $B_\perp$. The level width
contribution is indicated by systematic error bars, see text. The
solid line corresponds to an effective $g$ factor $g\approx
2.6$.}\vspace{-0.1in}
\end{figure}

Tilting the magnetic field allows us to verify the systematics of the
gaps in the spectrum and probe the lowest-energy charge-carrying
excitations. As the thickness of the 2D electron system in Si MOSFETs
is small compared to the magnetic length in accessible fields, the
parallel field couples largely to the electrons' spins while the
orbital effects are suppressed \cite{simonian}. Therefore, the
variation of a gap with $B_\parallel$ should reflect the change in
the excitation energy as the Zeeman splitting, $g\mu_BB$, is
increased: the excitation energy change is determined by the
difference between the spin projections onto magnetic field for the
ground and the lowest excited states. Within single-particle picture,
e.g., one can expect that with increasing $B_\parallel$ at fixed
$B_\perp$, the spin gap will increase, the valley gap will stay
constant, and the cyclotron gap, which is given by the difference
between the cyclotron splitting and the sum of the spin and the
valley splittings, will decrease. In contrast, for spin textures
(so-called skyrmions), the dependence of the excitation energy on
$B_\parallel$ should be much stronger compared to the single-particle
Zeeman splitting \cite{yang}.

\begin{figure}\vspace{2mm}
\scalebox{0.45}{\includegraphics[clip]{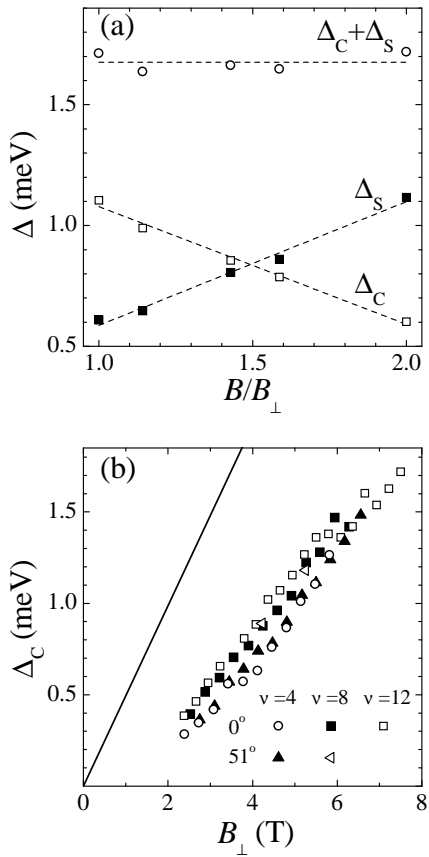}}
\caption{\label{fig3} (a) Change of the $\nu=6$ spin and the $\nu=4$
cyclotron gaps with $B_\parallel$ at fixed $B_\perp=5.5$~T. The
dashed lines are guides to the eye. (b) Chemical potential jump
across the cyclotron gap as a function of perpendicular field
component including the term that is responsible for the increase of
$\Delta_s$ with $B_\parallel$. Also shown by a solid line is the
difference between the single-particle cyclotron and Zeeman
splittings.}\vspace{-0.1in}
\end{figure}

In Fig.~\ref{fig2}(a), we show the value of the chemical potential
jump, $\Delta_s$, across the $\nu=2$ and $\nu=6$ gaps as a function
of magnetic field for different tilt angles. It is insensitive to
both filling factor and tilt angle, as expected for spin gaps. The
data are best described by a proportional increase of the gap with
the magnetic field with a slope corresponding to an effective $g$
factor $g\approx 1.75$. The so-determined value obviously gives a
lower boundary for the $g$ factor because both the valley splitting
at odd $\nu$ and the level width are disregarded.

In Fig.~\ref{fig2}(b), we show how the $\nu=6$ gap changes with
$B_\parallel$ at different values of the perpendicular field
component. It is noteworthy that the level width contribution, which
is indicated by systematic error bars, depends weakly on parallel
field, and the valley splitting has been verified to be independent
of $B_\parallel$. This, therefore, allows more accurate determination
of the $g$ factor as shown by the solid line in Fig.~\ref{fig2}(b).
Its slope yields $g\approx 2.6$, which is in agreement with the data
obtained for the $\nu=2$ and $\nu=10$ gaps. The fact that this value
is close to the $g$ factor $g=2$ in bulk silicon points to the single
spin-flip origin of the excitations for the $\nu=2$, 6, and 10 gaps.

Unlike spin gaps, the chemical potential jump, $\Delta_c$, across the
$\nu=4$, 8, and 12 gaps decreases with parallel magnetic field
component, as already seen from Fig.~\ref{fig1}. In
Fig.~\ref{fig3}(a), we compare the behaviors of the $\nu=6$ and
$\nu=4$ gaps with $B_\parallel$ at fixed perpendicular field
component. For $B_\perp$ between 2.7 and 6.6~T, the absolute values
of the slopes of these dependences are equal, within experimental
uncertainty, to each other so that the sum of the gaps is
approximately constant even if the level width contribution is taken
into account. These results lead to two important consequences: (i)
the $\nu=4$, 8, and 12 gaps are cyclotron ones, the conventional
systematics of the gaps remaining valid in the studied electron
density range down to $1.5\times 10^{11}$~cm$^{-2}$; and (ii) the $g$
factor does not vary with filling factor $\nu$. Although our value of
$g\approx 2.6$ is in agreement with the previously measured ones
\cite{englert,gm,us}, we do not confirm the conclusion on
oscillations of the $g$ factor with $\nu$ based on activation energy
measurements and made in line with theoretical predictions
\cite{ando,afs} under the assumption of $B_\parallel$-independent
level width \cite{englert}.

In Fig.~\ref{fig3}(b), we compare the data for the chemical potential
jump across the $\nu=4$, 8, and 12 gaps in perpendicular and tilted
magnetic fields including the term $g\mu_B(B-B_\perp)$ that describes
the increase of the spin gap with $B_\parallel$. The data coincidence
confirms that the changing spin gap is the only cause for the
dependence of the cyclotron gap on parallel field component. As is
evident from the figure, $\Delta_c$ is considerably smaller than the
value ($\hbar\omega_c-2\mu_BB_\perp$) expected within single-particle
approach ignoring both valley splitting and level width.

\begin{figure}\vspace{2mm}
\scalebox{0.45}{\includegraphics[clip]{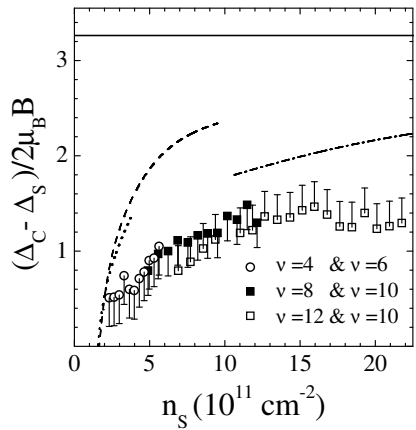}}
\caption{\label{fig4} Difference of the normalized values of the
cyclotron and the spin gaps in a perpendicular magnetic field vs
$n_s$. The resulting level width contribution is indicated by
systematic error bars. Also shown is the value ($m_e/m-g$) determined
from the data of Ref.~\cite{gm} (dashed line), Ref.~\cite{us} (dotted
line), and Ref.~\cite{smith72} (dash-dotted line) as well as using
the band electron mass and the $g$ factor in bulk silicon (solid
line).}\vspace{-0.1in}
\end{figure}

To reduce experimental uncertainty related to the inaccurate
determination of the level width, we plot in Fig.~\ref{fig4} the
difference, $(\Delta_c-\Delta_s)/2\mu_BB$, of the normalized values
of the cyclotron and the spin gaps in a perpendicular magnetic field
as a function of electron density. Assuming that the cyclotron
splitting is determined by the effective mass $m$, this difference
corresponds to ($m_e/m-g$), where $m_e$ is the free electron mass.
Using data for $m$ and $g$ obtained in both parallel \cite{gm} and
weak \cite{us,smith72} magnetic fields, we find that the value
($m_e/m-g$) is indeed consistent with our data, see Fig.~\ref{fig4}.
The effective mass determined from our high-$n_s$ data using $g=2.6$
is equal to $m\approx 0.23m_e$, which is close to the band mass of
$0.19m_e$. As long as our $g$ value is constant, the decrease of the
normalized gap difference with decreasing $n_s$ reflects the behavior
of the cyclotron splitting, which is in agreement with the conclusion
of the strongly enhanced effective mass at low electron densities
\cite{gm,us}.

We now discuss comparatively the results obtained for the valley and
the spin gaps. According to Ref.~\cite{valley}, the enhanced valley 
gap at the lowest filling factors $\nu=1$ and $\nu=3$ in Si MOSFETs 
is comparable to the single-particle Zeeman splitting. As our data 
for the spin gap correspond to the single-particle Zeeman splitting, 
this may lead to a different systematics of the gaps in the spectrum 
compared to the single-particle picture. Such a possibility has been 
supported by a recent theory \cite{brener} which shows the importance 
of the Jahn-Teller effect for the ground state of a 2D electron 
system in bivalley (100)-Si MOSFETs in quantizing magnetic fields. At 
$\nu=2$, due to static and dynamic lattice deformations, the valley 
degeneracy is predicted to lift off giving rise to a complicated 
phase diagram including three phases: spin-singlet, canted 
antiferromagnet, and ferromagnet. In our experiment, over the studied 
range of magnetic fields down to 3~T, we observe at $\nu=2$ a 
spin-ferromagnetic ground state only. This gives an estimate of the 
strength of the suggested mechanism for the valley splitting 
enhancement.

The fact that we do not observe oscillations of the $g$ factor as a
function of $\nu$ is not too surprising, because our value of $g$ is
close to the $g$ factor in bulk silicon so that those oscillations
may be small. At the same time, our data for the $g$ factor allow us
to arrive at a conclusion that at $\nu=2$, the valley gap is small
compared to the spin gap. Therefore, the valley splitting does
oscillate with filling factor \cite{ando}, the conclusion being
valid, at least, for the strongly enhanced gaps at $\nu=1$ and
$\nu=3$. We stress that this effect occurs in the strongly correlated
electron system, which is beyond the conventional theory of
exchange-enhanced gaps \cite{ando}.

Let us finally discuss the results obtained for the cyclotron gap.
The data of Fig.~\ref{fig4} indicate unequivocally that the origin of
the small $\Delta_c$ value in Fig.~\ref{fig3}(b) is not related to
valley splitting and level width. Instead, it is renormalization of
the effective mass and $g$ factor due to electron-electron
interactions: the observed decrease of the gap difference with
decreasing $n_s$ in Fig.~\ref{fig4} as well as the systematics of the
gaps are in agreement with both the decrease of the ratio of the
cyclotron and the spin gaps with decreasing $n_s$ \cite{krav00} and
the sharp increase of the effective mass at low electron densities
\cite{gm,us}. Needless to say that the conventional theory
\cite{ando} yields an opposite sign of the interaction effect on the
cyclotron splitting.

We gratefully acknowledge discussions with I.~L. Aleiner, S.~V.
Iordanskii, A. Kashuba, and S.~V. Kravchenko. This work was supported
by the RFBR, the Russian Ministry of Sciences, and the Programme
``The State Support of Leading Scientific Schools''. V.T.D.
acknowledges support of A.~von Humboldt foundation via
Forschungspreis.




\end{document}